\documentclass[bibyear]{aa} 

\usepackage{ulem}
\usepackage{graphicx}
\usepackage{natbib}
\usepackage[hidelinks,colorlinks=true,linkcolor=red,citecolor=blue]{hyperref}
\usepackage{txfonts}
\begin{document}

   \title{A horseshoe-shaped ring of diffuse emission detected at 1.4 GHz }
\titlerunning{A horseshoe-shaped ring of diffuse emission}

\author{Shobha Kumari
          \inst{1}
          \and
          Sabyasachi Pal\inst{1}
}
 \authorrunning{Kumari \& Pal}

   \institute{Midnapore City College, Kuturia, Bhadutala, Paschim Medinipur, West Bengal, 721129, India
             \thanks{E-mail: sabya.pal@gmail.com (SP)
             }
}

\def \src{J1407+0453}
 
  \abstract
  { 
  We identify a horseshoe-shaped ring (HSR) of diffuse emission in \src{} from the Faint Images of Radio Sky at Twenty-cm (FIRST) survey using the Very Large Array telescope at 1.4 GHz. An optical galaxy, SDSSJ140709.01+045302.1, is present near the limb of the HSR of \src{}, with a spectroscopic redshift of $z=0.13360$. The total extent of the source, including the diffuse emission of \src{}, is 65 arcsec (with a physical extent of 160 kpc), whereas the diameter of the HSR is approximately 10 arcsec (25 kpc). The flux density of the HSR is $\sim$47 mJy at 1400 MHz, whereas the flux densities of the whole diffuse emission of \src{} at 1400 MHz and 150 MHz are 172 mJy and 763 mJy, respectively. We measure the radio luminosity of HSR \src{} as 1.94 $\times 10^{24}$ W~Hz$^{-1}$, with a spectral index, $\alpha_{150}^{1400}=-0.67$. The black hole mass of \src{} is 5.8$\times10^8$ M$_{\odot}$. We compare the HSR of diffuse emission of \src{} with other discovered diffuse circular sources. The possible formation scenarios for \src{} are discussed, so as to understand the nature of the source. We present a spectral index map of source J1407+0453 to study the spectral properties of the source.
}
   
   \keywords{galaxies: ISM--galaxies: active--galaxies: clusters: intracluster medium--galaxies: kinematics and dynamics
               }
   \maketitle
\section{Introduction}
Diffuse radio emission are typically found in radio halos, radio relics, and mini-halos \citep{Fe96, Kem04}, depending on their location in the cluster and the type of cluster (merging or cool-core).
Clusters that show signs of merging processes are known to have radio halos at their centres \citep{Fe02, Gi02} that are an arcmin or more in diameter \citep{Fe21}. Relics are located close to the edges of both the merging and relaxed clusters. Mini-halos are centrally placed, hosted in relaxed cool-core clusters, and typically found around powerful radio galaxies. Mini-halos are diffuse radio emission that are typically a few 100 kpc in size. Diffuse extended emission in the Phoenix cluster \citep{va14}, Perseus cluster \citep{Si93}, RXCJ1720.1+2638 \citep{Gi14}, and very recently the Abell 1413 cluster \citep{Ri23} are examples of prototypes of the class of mini-halos. More commonly, the morphology of these diffuse sources has an irregular shape, but some radio relics also have more symmetric and roundish morphologies \citep[e.g. A1664, A2443,][]{Go01, Co11}.   
Here, it is worth noting that non-thermal diffuse emission can also be present outside the clusters, implying the existence of magnetic fields and relativistic particles in extremely low-density environments.

Apart from the diffuse emission associated with galaxy clusters, some roundish or horseshoe-shaped diffuse structures of about one arcmin or more in diameter have been observed in the simulation study of \citet{Do23}. Recently a circular symmetry extended diffuse radio emission (J1507+3013) around an elliptical galaxy was identified using the Very Large Array (VLA) Faint Images of Radio Sky at Twenty-cm (FIRST) survey \citep{Ku23}. The emergence of such diffuse emission, by taking a virial mass of 10$^{12}$ M$_{\odot}$, may be favoured by merger-driven internal shocks, as is suggested by \citet{Do23}. The gravitationally hierarchical emergence of large-scale structures causes shocks in the intergalactic medium (IGM). Shocks can either re-accelerate prior relativistic electron populations produced by former active galactic nuclei (AGN) activity within a cluster or directly accelerate IGM thermal electrons \citep{Pi08}. The discovery of radio emission from intergalactic shocks will have a significant impact on our understanding of cosmology and astrophysics because it will test theories about how structures form, confirm the existence of the previously unknown warm-hot intergalactic medium, and map its distribution \citep{Ke04, Pi08}. 
 
Recently, very faint, edge-brightened diffuse radio emission, popularly known as odd radio circles (ORCs), have been identified using the pilot survey of the Evolutionary Map of the Universe \citep[EMU;][]{No21a}. Using this survey, three ORCs (ORC1 = ORC J2103--6200, ORC J2058--5736 = ORC2, and ORC J2059--5736 = ORC3) have been discovered at high galactic latitudes \citep{No21b}.
Using Giant Metrewave Radio Telescope (GMRT) archival data of \citet{ve17} (Proposal ID: 23\_017) at 325 MHz (discussed in \citet{No21b}), Australian Square Kilometre Array Pathfinder (ASKAP) data at 944 MHz \citep{Ko21}, and LOFAR Two-metre Sky Survey DR2 (LoTSS DR2) at 144 MHz \citep{Om22c}, three more ORCs have been discovered.
Previously, \citet{No21b, No21c, Ko21, No22, Om22a, Om22b} discussed various scenarios for the formation of ORCs, including the spherical shock wave from the optical host galaxy, termination shock from the starburst wind, and interactions between the optical galaxies inside the ring of the structure.
\begin{figure*}
\vbox{
\centerline{
\includegraphics[width=9.3cm, height=8.0cm,origin=c]{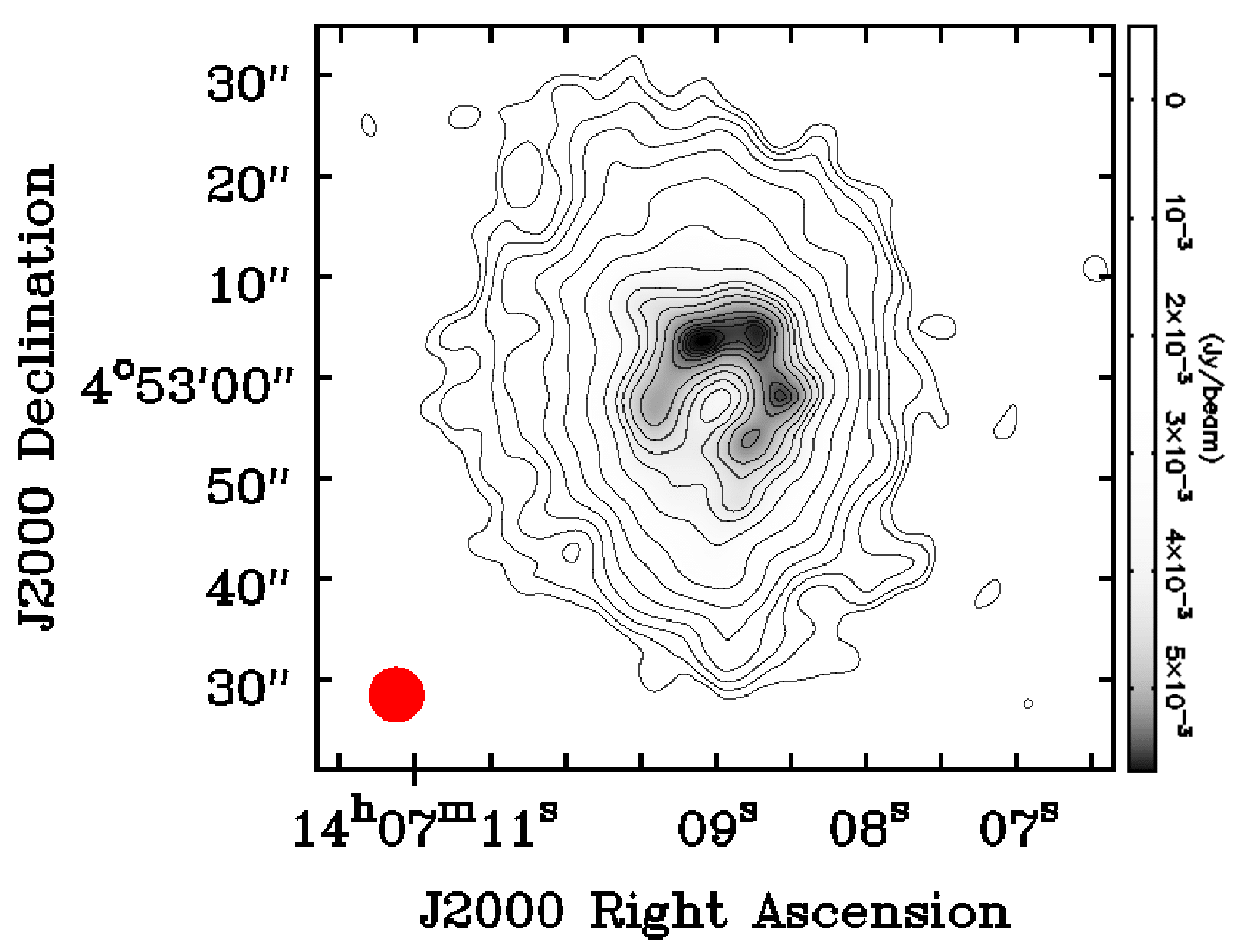}
\includegraphics[width=9.3cm, height=8.0cm,origin=c]{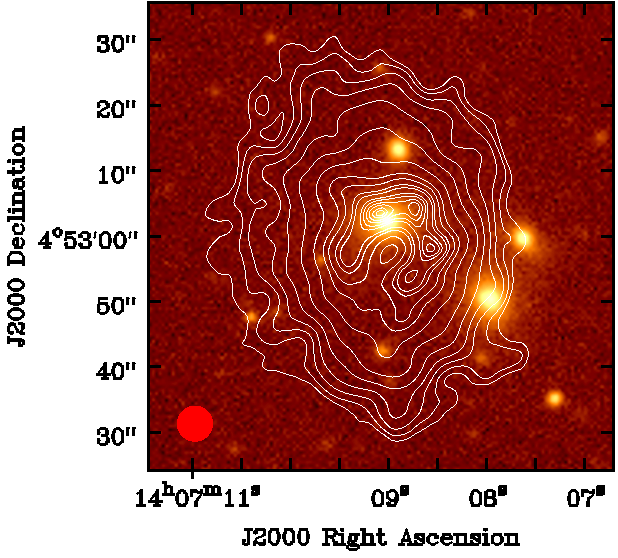}
}
}
\caption{Left: \src{} at 1400 MHz. The grey scale represents the flux density of the source. Right: \src{} at 1400 MHz in a contour overlaid with the optical image taken from the 9th data release of the Dark Energy Spectroscopic Instrument's Legacy Imaging Surveys \citep[DESI LS DR9;][]{Sc21}. For both images, the contour levels are at 3$\sigma\times$(1, 1.4, 2, 2.8, 4, 5.7, 8, 10, 11, 12, 12.5, 13.2, 13.5, 13.8, 14, 14.2, 14.3, 15), where $\sigma$ = 133 $\mu$Jy beam$^{-1}$ is the mean rms noise around the FIRST map.}
\label{fig:orc0}
\end{figure*}

The current paper discusses a newly identified horseshoe-shaped ring (HSR) of diffuse emission. The source identification is described in Sect. \ref{sec:identi}. In Sect. \ref{sec:results}, we present our results. In Sect. \ref{sec:discuss}, we compare our results with previously known similar sources and discuss the possible formation scenarios of \src{}. Section \ref{sec:conclusion} summarises the results of this study. We used the following $\Lambda$CDM cosmology parameters for the current paper using the final full-mission Planck measurements of the CMB anisotropies \citep{Ag20}: $H_0$ = 67.4 km s$^{-1}$ Mpc$^{-1}$, $\Omega_{vac}$ = 0.685, and $\Omega_m$ = 0.315.

\section{A horseshoe-shaped ring of diffuse emission}
\label{sec:identi}
We report a newly discovered HSR of diffuse emission in \src{} from the Faint Images of Radio Sky at Twenty-cm (FIRST) survey at 1.4 GHz using the Very Large Array (VLA) radio telescope \citep{Be95, Wh97}. 
An optical galaxy is present near the limb of the horseshoe-shaped inner ring structure of \src{}, which may be the host of the diffuse source. The optical galaxy of HSR \src{} has a spectroscopic redshift of 0.13360, obtained from SDSS \citep{Ahu20}. The total angular diameter of \src{} is 65 arcsec, with a physical extent of approximately $\sim$160 kpc, whereas the HSR has an angular diameter of 10 arcsec with a physical extent of $\sim$ 25 kpc. The source diameter was measured by fitting a circle onto the source. Figure \ref{fig:orc0} shows the FIRST image of the HSR \src{} at 1400 MHz.

The source presented in the current paper (J1407+0453) has also been detected in other surveys, such as the Tata Institute of Fundamental Research (TIFR) Giant Metre-wave Radio Telescope (GMRT) Sky Survey (TGSS) at 150 MHz \citep{In17}, the National Radio Astronomy Observatory (NRAO) VLA Sky Survey (NVSS) at 1400 MHz \citep{Co98}, the VLA Sky Survey (VLASS) at 3 GHz \citep{La20}, the Galactic and Extragalactic All-sky Murchison Widefield Array (GLEAM) at 72--231 MHz \citep{Hu17}, and The Rapid ASKAP Continuum Survey (RACS) at 888 MHz \citep{Mc20}. All these surveys detected J1407+0453 as a point source, except for VLASS, which shows the diffuse emission structure of the source.

\section{Results}
\label{sec:results}
\subsection{Radio Properties of \src{}}
\label{sec:radio}
We measure the spectral index of \src{} between 1400 MHz and 150 MHz, using $F_{\nu} \propto \nu^{\alpha}$, where $F_{\nu}$ is the integrated flux density at frequency $\nu$ and $\alpha$ is the spectral index. The spectral index for the total diffuse emission of \src{} is calculated as $\alpha_{150}^{1400}$= --0.67$\pm0.05$ using integrated radio flux densities of 172 and 763 mJy at 1400 MHz and 150 MHz, respectively, of the total diffuse emission of \src{}. Here, we use the NVSS \citep{Co98} flux density instead of the FIRST because the FIRST survey is prone to flux density loss owing to the lack of short spacing (short baselines). 
The radio luminosity of the total diffuse emission of \src{} at 1400 MHz is measured as 7.09$\times$10$^{24}$ W Hz$^{-1}$ with the help of the standard formula \citep{Do09}\begin{equation}
    L_{\textrm{rad}}=4\pi{D_{L}}^{2}F_{\nu}(1+z)^{\alpha-1}
,\end{equation}
where $F_{\nu}$ is the flux density (W m$^{-2}$ Hz$^{-1}$) at a given frequency, $D_{L}$ is the luminosity distance to the source in metres (m), $\alpha$ is the spectral index, and $z$ is the redshift of \src{} ($z=0.13360$).
The flux density of the HSR of \src{} at 1400 MHz is measured as $\sim$47 mJy. Using the flux density of the HSR, we calculate the radio luminosity of the HSR of \src{} to be 1.94$\times$ 10$^{24}$ W H$z^{-1}$.

\begin{figure}
\vbox{
\centerline{
\includegraphics[width=9.3cm, height=7.3cm,origin=c]{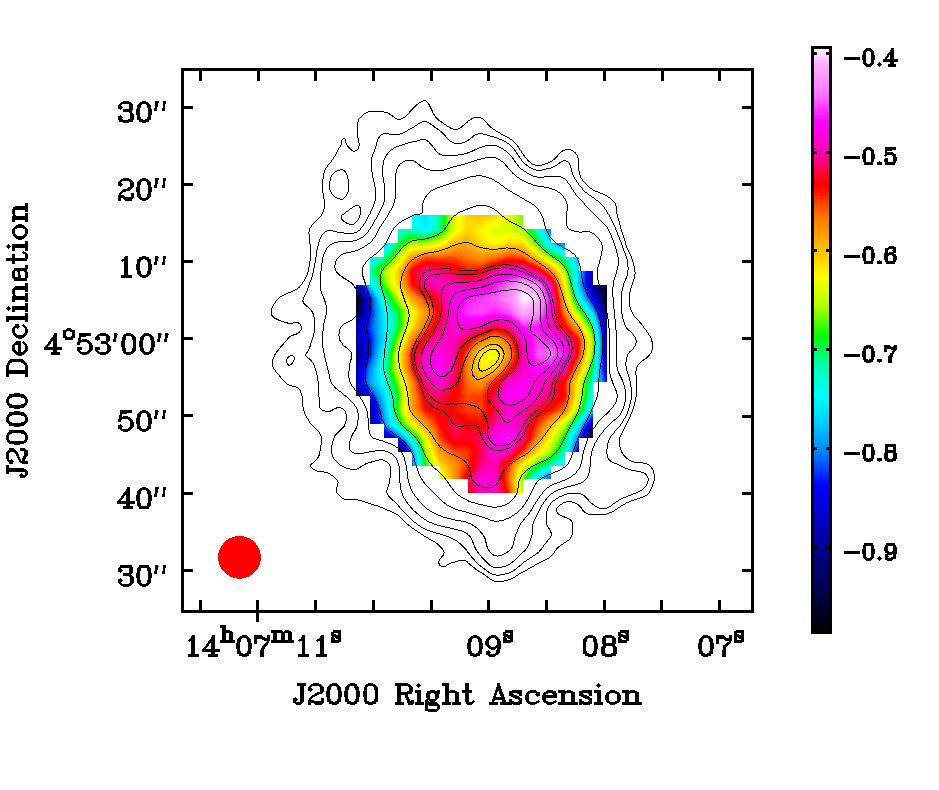}
}
}
\caption{Spectral index map of \src{} overlaid with the FIRST image in contour. The lowest contour level in the figure is at 3$\sigma$, where $\sigma$ is the rms noise of the source in the FIRST map. The rms noise of the source in the VLA FIRST image is 133 $\mu$Jy beam$^{-1}$. The contour levels are at 3$\sigma\times$(1, 1.4, 2, 2.8, 4, 5.7, 8, 10, 11, 12, 13, 14, 15). The colour scale represents the spectral index of \src{}.}
\label{fig:spec_map}
\end{figure}

\subsection{Spectral index map of \src{}}
\label{sec:spec_map}
We present the spectral index map of \src{} in Fig. \ref{fig:spec_map} between 1400 MHz and 150 MHz.
The typical error in the spectral index map is $\sim 0.02$. The error is nearly uniform, except at the edge of the radio emission, where it reaches 0.08. The contour plot in Fig. \ref{fig:spec_map} shows the VLA FIRST image of \src{} at 1400 MHz. The spectral index of HSR ranges from --0.4 to --0.5. The spectral index inside the ring is --0.65, whereas the spectral index of the outer region of the HSR becomes steeper, ranging from --0.55 to --0.95. Spectral steepening from the core towards the edge of the lobes is observed in radio galaxies, which is thought to be due to spectral ageing in the structure of the source. The steepening in the outer region of the HSR is also possibly due to spectral ageing in the structure of the source. 
 
\subsection{Optical properties of \src{}}
\label{sec:optical}
We identify the optical counterpart of \src{} from the Dark Energy Spectroscopic Instrument's Legacy Imaging Surveys \citep[DESI LS DR9;][]{Sc21}. An overlaid image of \src{} at 1.4 GHz with DESI is shown in the right panel of Fig. \ref{fig:orc0}. The optical galaxy SDSS J140709.01+045302.1 resides near the geometrical centre of \src{} (labelled as the “core” in Fig. \ref{fig:optical}), with a spectroscopic redshift of 0.13360. The location of SDSS J140709.01+045302.1 also coincides with the radio core of \src{}. Within a distance of one arc minute from the optical galaxy SDSSJ140709.01+045302.1, there are six optical or infrared (IR) sources labelled as “A”, “B”, “C”, “D”, “E”, and “F” (see Fig. \ref{fig:optical}). The bright optical or IR source labelled as “A” (2MASS 140708.914+045313.48) is at an angular distance of 11 arcsec (31 kpc projected distance) from the optical galaxy SDSS J140709.01+045302.1. The photometric redshift of the optical galaxy A is 0.158$\pm0.021$. 
A galaxy labelled as B (see Fig. \ref{fig:optical}) is located to the southeast of A and at an angular distance of 20.86 arcsec (54 kpc projected distance) from the optical galaxy SDSSJ140709.01+045302.1. This galaxy is 2MASS J14070763+0452592 and has a photometric redshift of 0.141$\pm0.012$. An optical galaxy labelled as C (see Fig. \ref{fig:optical}) is located to the southwest of B and at 19 arcsec angular distance (44 kpc projected distance) from the optical galaxy SDSSJ140709.01+045302.1. This galaxy is WISEA J140707.95+045250.5, or 2MASS J14070795+0452505, and has a photometric redshift of 0.125$\pm0.008$. Three more galaxies labelled as D, E, and F (see Fig. \ref{fig:optical}) are located at 19.92 arcsec (78.3 kpc projected distance), 25.54 arcsec (51.51 kpc projected distance), and 23.62 arcsec (116.5 kpc projected distance) from the optical galaxy SDSS J140709.01+045302.1, respectively. The photometric redshifts of galaxies D, E, and F are 0.240$\pm$0.071, 0.106$\pm0.118$, and 0.331$\pm0.190$, respectively. The optical galaxy D is 2MASS J140709098+045242.27, E is 2MASS J140710446+045248, and F is WISEA J140709.10+045325.3. Photometric redshifts are taken from the Dark Energy Camera Legacy Survey data release 9 \citep[DECaLS DR9;][]{De19} and a spectroscopic redshift of SDSS J140709.01+045302.1 (labelled as the “core” in Fig. \ref{fig:orc0}) is taken from SDSS \citep{Ahu20}.

We measure the approximate black hole mass of the optical host galaxy (SDSSJ140709.01+045302.1) of \src{} using the $M_{BH}$--$\sigma_{\ast}$ relation \citep{Ge00,Fe00} \begin{equation}
        \frac{M_{BH}}{10^8M_\odot}=3.1 \left(\frac{\sigma_\ast}{200 \text{~km~s}^{-1}}\right)^4 
,\end{equation}
where $\sigma_{\ast}$ is the velocity dispersion and $M_{\odot}$ is the stellar mass ($\sim$2 $\times$$10^{31}$ kg). We used the stellar velocity dispersion ($\sigma_{\ast}$) information from the SDSS \citep{Ahu20} to compute $M_{BH}$. The black hole mass of \src{} is measured as 5.8 $\times10^8$ $M_{\odot}$ ($\sim$10$^{8.8}$ $M_{\odot}$) with a stellar velocity dispersion ($\sigma_{\ast}$) of 233.46$\pm${8.70} km/s. 

\subsection{Mid-IR properties of \src{}}
To study the host properties of \src{} at mid-IR wavelengths, we used the Wide-field Infrared Survey Explorer \citep[WISE;][]{Wr10}, which is an all-sky IR survey. WISE is a space-based telescope that can observe in four bands (W1, W2, W3, and W4) corresponding to 3.4 $\mu$m, 4.6 $\mu$m, 12 $\mu$m, and 22 $\mu$m wavelengths, respectively. We measured mid-IR colours W1--W2 = 0.13, and W2--W3 = 1.24, which suggests that \src{} may be a low-excitation radio galaxy (LERG) \citep{Pr18, Da20}. It should be noted that WISE data alone are not sufficient to classify the host galaxy of \src{}; we need a detailed optical study to understand the nature of the host galaxy of \src{}.
 
\section{Discussions}
\label{sec:discuss}
\subsection{HSR \src{} versus previously discovered diffuse sources}
\label{subsec:comp}
Nearly circular-shaped diffuse emission (for example, Fig 8a and 8b of \citet{Sa02}) is not uncommon, but a ring-like structure within a circular-shaped diffuse emission is rare. Recently discovered ORCs also appear as ring-like structures with a bright edge, along with an interior that is either faint or devoid of radio emission. Among the six previously discovered ORCs, three are associated with optical galaxies near their geometrical core, and the diffuse source, \src{}, in the current study is also associated with an optical galaxy near the core.
Here, we compare the morphological features and properties of \src{} with those of three previously discovered, optically hosted ORCs (ORC J2103--6200 (ORC1), ORC J1555+2726 (ORC4), and ORC J0102--2450 (ORC5)).

The inner edge of \src{} is bent with three bright compact radio counterparts, making it an HSR (see left and right panels of Fig. \ref{fig:orc0}), whereas the outer edge of the ring of ORC5 is bent (like a wide-angle tailed (WAT) radio galaxy). The similarity between \src{} and ORCs 1, 4, and 5 is that all of the sources have a radio core inside the ring, but the possible optical galaxy resides at the limb of the HSR in \src{} and not in the radio core (see right panel of Fig. \ref{fig:orc0}), whereas in ORCs 1, 4, and 5 the optical counterpart coincides with the radio core. Initially, only one outer ring was observed in ORC1, but in the MeerKAT observation, apart from the outer ring, an arc-like structure inside the ring (also referred to as a polar ring) was observed \citep{No22}, whereas \src{} contains only the horseshoe-shaped inner ring without an outer ring.

The WISE colours for \src{} (W1--W2 = 0.13, W2--W3 = 1.24) and ORC1 (W1--W2 = 0.081, W2--W3 < 2.045) for the central object suggest that both of the sources are low-excitation radio galaxies (LERGs) (W1--W2 < 0.5 and 0.2 < W2--W3 < 4.5) \citep{Pr18, Da20}. The HSR of diffuse emission of \src{} has a flux density of $\sim$47 mJy at 1400 MHz in the FIRST image, whereas the total diffusion emission of \src{} has radio flux densities of 172 mJy and 763 mJy at 1400 MHz and 150 MHz, respectively. Therefore, \src{} has a significantly higher flux density than ORC1, ORC4, and ORC5, which have flux densities in the range of 1.9 mJy to 7 mJy at $\sim$1 GHz. The spectral index of the HSR of diffuse emission of \src{} ranges from --0.4 to --0.5, whereas the outer region of the HSR has a spectral index in the range of --0.55 to --0.95. So, \src{} has a flatter radio spectrum in the periphery of the HSR and along the outer region of the HSR compared to the spectral indices of ORC1, ORC4, and ORC5, as they have a very steep spectral index $\leq$--0.9 \citep{No21b, No21c, No22, Ko21}. 

The physical extent of the HSR ring of \src{} is $\sim$25 kpc with a spectroscopic redshift of 0.13360, whereas the physical extents of ORC1, ORC4, and ORC5 lie in the range of 100--450 kpc (with available photometric redshifts in the range of 0.39--0.55).  

The black hole mass of \src{} is 5.8 $\times10^8$ M$_{\odot}$, whereas that of ORC5 is 7.5$\times10^8$ M$_{\odot}$. The radio luminosity of the HSR of \src{} is 1.94$\times$ 10$^{24}$ W H$z^{-1}$ and that of the total diffuse emission of \src{} is 7.09$\times$ 10$^{24}$ W H$z^{-1}$, whereas for ORC1, ORC4, and ORC5, the radio luminosity is $\sim$10$^{23}$ W H$z^{-1}$. Thus, the black hole in \src{} is less massive, but the ring is more luminous than those of ORC1, ORC4, and ORC5. 

 \begin{figure}
\includegraphics[width=9cm, origin=c]{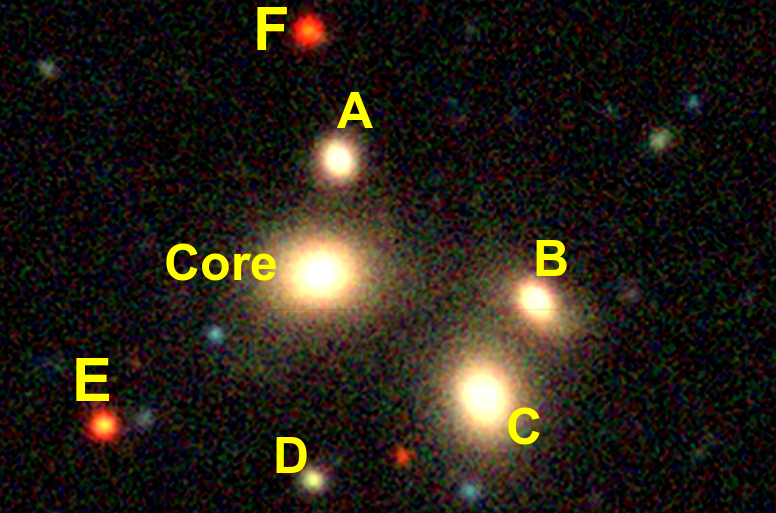}
\caption{Optical image of \src{} region taken from DESI LS DR9 \citep{Sc21}.}
\label{fig:optical}
\end{figure}

Morphologically, \src{} and the previously discovered ORCs have similarities, with a ring of diffuse emission. However, as we compare \src{} with ORC1, ORC4, and ORC5 above, \src{} is more luminous, with a significantly strong flux density and a less steep spectrum.
\src{} also exhibits a horseshoe-shaped inner ring about $\sim$25 kpc in size (much smaller than the ring observed in previously discovered ORCs; 100--450 kpc), and none of the ORCs exhibit such a horseshoe-shaped inner ring. Although the HSR of \src{} is much smaller than that of the ORCs, the size of the total diffuse emission of \src{} is $\sim$160 kpc, which is comparable in size to that of the ORCs. However, we should be careful about the fact that there are only three ORCs known so far that are hosted by optical galaxies.
More sources with similar properties are required to understand the nature of ORCs along with \src{}. It is also possible that \src{} may be in its early phase, yet to lose its energy and become a low-power faint ORC, similar to other discovered ORCs.

Diffuse radio emission have also been observed in galaxy clusters such as mini-halos \citep{Kem04}. Mini-halos are often seen around radio galaxies within clusters. It should be noted that these mini-halos are significantly larger ($\geq$100 kpc in size) \citep{Ri23} than \src{}. It is also difficult to classify \src{} as a mini-halo because \src{} is not associated with any galaxy cluster.

Another possibility is that the diffuse emission of \src{} may represent the remains of a “fossil” radio galaxy \citep{Kem04, Ri23}, but that would be expected to have a very steep spectral index (<--1.5), which is not supported by the results presented in the current paper for \src{}.

\subsection{Formation scenarios of \src{}}
\label{subsec:forma}
In the following, we discuss some plausible formation scenarios for \src{}. 
One of the possibilities for the formation of \src{} could be the shocks produced in post-starburst galaxies by tidal disruption events around the core of supermassive black holes (as was previously discussed for the formation of circular diffuse sources like ORCs) \citep{Om22a, Om22b}. This could also be considered for diffuse sources with optical host galaxies near their geometrical centre, as is observed in \src{}. 

Another possible hypothesis is that \src{} could be the result of a giant blast wave in the central galaxy \citep{No21b, No21c, No22, Ko21}. This blast wave produces an HSR of diffuse radio emission that appears as edge-brightened discs similar to planetary nebulae or supernova remnants (SNRs). The radio emission is likely produced due to the synchrotron emission caused by accelerated electrons as a result of the shock \citep{Do02, Ca12, Ri21}.
A binary SMBH merger in the centre of the galaxy could produce such a spherical shock \citep{Bo12}; thus, we would expect to observe a magnetic field in the ring that is largely tangential and orthogonal to the velocity of the shock, such as SNRs or cluster relics. 

Recently, a simulation study of \citet{Do23} explained the internal merger shocks produced around some galaxies by matching several observed properties of the neighbouring galaxy. This simulation study shows that shocks produced by strong galactic merger events cause circular radio emission (structure) with a virial mass of 10$^{13}$ M$_\odot$ \citep{Do23}. In their simulation, a major merger event at $z$ = 0.7 is possibly responsible for driving two subsequent shocks far beyond the virial radius propagation \citep{Do23}. In this propagation, roundish- or horseshoe-shaped, roughly arcmin-size structures of diffuse emission are produced \citep{Do23}, similar to \src{} presented in the current paper. Noteworthily, \citet{Do23} discovered radio rings that are far fainter and larger than the HSR observed in \src{}.

There are six nearby optical sources within an angular diameter of 1 arcmin from the optical core of \src{} (as is discussed in Sect. \ref{sec:optical}). The nearest optical galaxies are A (photometric redshift of 0.158$\pm$0.021) and C (photometric redshift of 0.125$\pm$0.008), with projected distances from the optical host galaxy of 31 and 44 kpc, respectively. The presented uncertainties on the redshifts for optical galaxies A, B, C, D, E, and F (see Fig. \ref{fig:optical}) suggest that these galaxies are several Mpc from each other. However, uncertainties are often underestimated, and thus galaxies may be associated. Therefore, it is possible that \src{} is the result of the production of shock waves that led to the ring-like diffuse structure of \src{}, as is discussed above and observed in \citet{Do23}.  

Here, it should be noted that the elliptical host galaxy of \src{} is located on the limb of the horseshoe shape rather than in the centre. Therefore, this morphology may not be consistent with the emission of a spherical shockwave from an elliptical optical galaxy.

The horseshoe-shaped inner ring of \src{} resembles the morphology of narrow angle-tailed (NAT) sources \citep{Mi19, Bh22, Sa22, Pa23}. However, most of the previously discovered NAT sources are associated with a cluster of galaxies, and for \src{} no associated galaxy cluster is identified. For NAT sources, the optical counterparts are located near the centre of the bent structure between the two warm spots \citep{Bh22}. However, the HSR of \src{} does not contain any symmetrical warm spots and the optical counterpart in \src{} is offset from the centre. The optical counterpart is neither inside the ring nor located on the periphery of the ring; instead, the optical counterpart is located on the limb of the HSR. 
Therefore, \src{} may not be a head-tailed source.

As \src{} is unlikely to fall in the above-discussed classes of diffuse sources, it may represent a new class of strong and luminous extended diffuse sources hosted by an optical galaxy near the centre of the structure. 

\section{Conclusions}
\label{sec:conclusion}
A diffuse source, \src{}, with a horseshoe-shaped inner ring is reported in this paper from the VLA FIRST survey at 1.4 GHz. The total angular diameter of \src{} is 65 arcsec, whereas that of the HSR is $\sim$10 arcsec (which is smaller than that of the previously known ring of ORCs). \src{} is associated with an optically dense region with a bright compact optical host galaxy identified near the limb of the HSR and six nearby optical galaxies within an angular distance of 1 arcmin from the core of the optical host galaxy. \src{} could be the early, young phase of an ORC object hosted by an optical galaxy or it can be a new class of object that needs to be studied in detail with more samples. The structure of such diffuse emission, as is presented in the current paper in \src{}, can also be seen in the class of mini-halos and fossil radio galaxies. \src{} may not be a mini-halo or fossil radio galaxy because \src{} is not associated with any galaxy cluster (required for mini-halo sources), and nor does it possess a steep spectrum (<--1.5) (needed for fossil radio galaxies). In this paper, we discuss some possible scenarios for the formation of \src{}.

Further follow-up multi-wavelength observations are encouraged to study the nature of the diffuse emission of \src{}. Deep X-ray observations and studies are also required to detect any energetic events inside the HSR of \src{} that may cause the diffuse emission in the structure.

\begin{acknowledgements} 
We thank the anonymous reviewer for helpful suggestions. This paper used Sloan Digital Sky Survey V data. Funding for the Sloan Digital Sky Survey V has been provided by the Alfred P. Sloan Foundation, the Heising-Simons Foundation, the National Science Foundation, and the Participating Institutions. SDSS acknowledges support and resources from the Center for High-Performance Computing at the University of Utah. The photometric redshifts for the Legacy Surveys (PRLS) catalogue are used in this paper. The PRLS is funded by the U.S. Department of Energy Office of Science, Office of High Energy
Physics via grant DE-SC0007914.
\end{acknowledgements}


\end{document}